\input harvmac
\input epsf
\newcount\figno
\figno=0
\def\fig#1#2#3{
\par\begingroup\parindent=0pt\leftskip=1cm\rightskip=1cm\parindent=0pt
\baselineskip=11pt \global\advance\figno by 1 \midinsert
\epsfxsize=#3 \centerline{\epsfbox{#2}} \vskip 12pt {\bf Fig.\
\the\figno: } #1\par
\endinsert\endgroup\par
}
\def\figlabel#1{\xdef#1{\the\figno}}
\def\encadremath#1{\vbox{\hrule\hbox{\vrule\kern8pt\vbox{\kern8pt
\hbox{$\displaystyle #1$}\kern8pt}
\kern8pt\vrule}\hrule}}
\baselineskip14pt

\def\ads{AdS$_3$}
\def\eps{$i\epsilon$}

\Title{\vbox{\baselineskip12pt
\hbox{hep-th/0212277}
\hbox{UCLA-02-TEP-41}
\hbox{CALT 68-2421}
\hbox{SU-ITP-02-45}
\vskip-.5in}}
{\vbox{\centerline{Inside the Horizon with AdS/CFT }}}

\medskip\bigskip
\centerline{Per Kraus$^1$, Hirosi Ooguri$^2$, and Stephen
Shenker$^{3}$}
\bigskip\medskip
\centerline{\it $^1$Department of Physics and Astronomy
UCLA, Los Angeles, CA 90095, U.S.A.}
\medskip
\centerline{\it $^2$California Institute of Technology 452-48,
Pasadena, CA 91125, USA}
\medskip
\centerline{\it $^3$Department of Physics, Stanford University,
Stanford, CA, 94305, USA}
\medskip\bigskip\medskip\bigskip\medskip
\baselineskip14pt

\noindent

Using the eternal BTZ black hole as a concrete example, we show how
spacelike singularities and horizons can be described in terms of
AdS/CFT amplitudes.  Our approach is based on analytically continuing
amplitudes defined in Euclidean signature.  This procedure yields
finite Lorentzian amplitudes.
The naive divergences associated with the
Milne type singularity of BTZ are regulated by an $i\epsilon$
prescription inherent
in the analytic continuation and a cancellation between future and
past
singularities.

The boundary description corresponds to a tensor product of two
CFTs in an entangled state, as in previous work.  We give two bulk
descriptions corresponding to two different analytic
continuations.  In the first, only regions outside the horizon
appear explicitly, and so amplitudes are manifestly finite.  In
the second, regions behind the horizon and on both sides of the
singularity appear, thus yielding finite amplitudes for virtual
particles propagating through the black hole singularity.  This
equivalence between descriptions only outside and both inside and
outside the horizon is reminiscent of the ideas of black hole
complementarity.

\Date{December, 2002}

\lref\Inami{
T.~Inami and H.~Ooguri,
Prog.\ Theor.\ Phys.\  {\bf 73}, 1051 (1985).
}

\lref\Burgess{
C.~P.~Burgess and C.~A.~Lutken,
Phys.\ Lett.\ B {\bf 153}, 137 (1985).
}

\lref\MaldacenaHW{
J.~M.~Maldacena and H.~Ooguri,
J.\ Math.\ Phys.\  {\bf 42}, 2929 (2001) [arXiv:hep-th/0001053];
Phys.\ Rev.\ D {\bf 65}, 106006 (2002) [arXiv:hep-th/0111180];
J.~M.~Maldacena, H.~Ooguri and J.~Son,
J.\ Math.\ Phys.\  {\bf 42}, 2961 (2001) [arXiv:hep-th/0005183];
J.~M.~Maldacena and H.~Ooguri,
functions,''
Phys.\ Rev.\ D {\bf 65}, 106006 (2002)
[arXiv:hep-th/0111180].
}

\lref\Ben{M. Berkooz, B. Craps, D. Kutasov, and G. Rajesh,
arXiv:hep-th/0212215.}

\lref\Tolley{ A.~J.~Tolley and N.~Turok,
Phys.\ Rev.\ D {\bf 66}, 106005 (2002) [arXiv:hep-th/0204091].
}

\lref\Herzog{ C.~P.~Herzog and D.~T.~Son,
arXiv:hep-th/0212072.
}


\lref\KutasovXU{
D.~Kutasov and N.~Seiberg,
JHEP {\bf 9904}, 008 (1999)
[arXiv:hep-th/9903219].
}

\lref\deBoerPP{
J.~de Boer, H.~Ooguri, H.~Robins and J.~Tannenhauser,
JHEP {\bf 9812}, 026 (1998)
[arXiv:hep-th/9812046].
}


\lref\CornalbaFI{ L.~Cornalba and M.~S.~Costa,
Phys.\ Rev.\ D {\bf 66}, 066001 (2002) [arXiv:hep-th/0203031].
}

\lref\CruzIR{
N.~Cruz, C.~Martinez and L.~Pena,
[arXiv:gr-qc/9401025].
}

\lref\MartinecXQ{
E.~J.~Martinec and W.~McElgin,
arXiv:hep-th/0206175.
}

\lref\GiveonNS{
A.~Giveon, D.~Kutasov and N.~Seiberg,
Adv.\ Theor.\ Math.\ Phys.\  {\bf 2}, 733 (1998)
[arXiv:hep-th/9806194].
}

\lref\SonQM{
J.~Son,
arXiv:hep-th/0107131.
}

\lref\HananyEV{
A.~Hanany, N.~Prezas and J.~Troost,
JHEP {\bf 0204}, 014 (2002)
[arXiv:hep-th/0202129].
}

\lref\TeschnerFT{
J.~Teschner,
Nucl.\ Phys.\ B {\bf 546}, 390 (1999) [arXiv:hep-th/9712256];
%
Nucl.\ Phys.\ B {\bf 571}, 555 (2000)
[arXiv:hep-th/9906215].
}

\lref\DanielssonZT{
U.~H.~Danielsson, E.~Keski-Vakkuri and M.~Kruczenski,
Nucl.\ Phys.\ B {\bf 563}, 279 (1999)
[arXiv:hep-th/9905227].
}
\lref\EvansFR{
T.~S.~Evans, A.~Gomez Nicola, R.~J.~Rivers and D.~A.~Steer,
arXiv:hep-th/0204166.
}

\lref\BanadosWN{
M.~Banados, C.~Teitelboim and J.~Zanelli,
Phys.\ Rev.\ Lett.\  {\bf 69}, 1849 (1992)
[arXiv:hep-th/9204099];
M.~Banados, M.~Henneaux, C.~Teitelboim and J.~Zanelli,
Phys.\ Rev.\ D {\bf 48}, 1506 (1993)
[arXiv:gr-qc/9302012].}

\lref\IsraelUR{
W.~Israel,
Phys.\ Lett.\ A {\bf 57}, 107 (1976).
}

\lref\UnruhDB{
W.~G.~Unruh,
Phys.\ Rev.\ D {\bf 14}, 870 (1976).
}

\lref\ElitzurRT{
S.~Elitzur, A.~Giveon, D.~Kutasov and E.~Rabinovici,
JHEP {\bf 0206}, 017 (2002)
[arXiv:hep-th/0204189].
}

\lref\CrapsII{
B.~Craps, D.~Kutasov and G.~Rajesh,
JHEP {\bf 0206}, 053 (2002)
[arXiv:hep-th/0205101].
}

\lref\CornalbaNV{
L.~Cornalba, M.~S.~Costa and C.~Kounnas,
arXiv:hep-th/0204261.
}

\lref\GubserBC{
S.~S.~Gubser, I.~R.~Klebanov and A.~M.~Polyakov,
Phys.\ Lett.\ B {\bf 428}, 105 (1998)
[arXiv:hep-th/9802109].
}

\lref\KeskiVakkuriNW{
E.~Keski-Vakkuri,
Phys.\ Rev.\ D {\bf 59}, 104001 (1999)
[arXiv:hep-th/9808037].
}

\lref\MaldacenaRE{
J.~M.~Maldacena,
Adv.\ Theor.\ Math.\ Phys.\  {\bf 2}, 231 (1998)
[Int.\ J.\ Theor.\ Phys.\  {\bf 38}, 1113 (1999)]
[arXiv:hep-th/9711200].
}

\lref\LifschytzEB{
G.~Lifschytz and M.~Ortiz,
Phys.\ Rev.\ D {\bf 49}, 1929 (1994)
[arXiv:gr-qc/9310008].
}

\lref\BalasubramanianSN{
V.~Balasubramanian, P.~Kraus and A.~E.~Lawrence,
Phys.\ Rev.\ D {\bf 59}, 046003 (1999)
[arXiv:hep-th/9805171].
}

\lref\WittenQJ{
E.~Witten,
Adv.\ Theor.\ Math.\ Phys.\  {\bf 2}, 253 (1998)
[arXiv:hep-th/9802150].
}

\lref\BalasubramanianRE{
V.~Balasubramanian and P.~Kraus,
Commun.\ Math.\ Phys.\  {\bf 208}, 413 (1999)
[arXiv:hep-th/9902121].
}

\lref\HorowitzXK{
G.~T.~Horowitz and D.~Marolf,
JHEP {\bf 9807}, 014 (1998)
[arXiv:hep-th/9805207].
}

\lref\BalasubramanianDE{
V.~Balasubramanian, P.~Kraus, A.~E.~Lawrence and S.~P.~Trivedi,
Phys.\ Rev.\ D {\bf 59}, 104021 (1999)
[arXiv:hep-th/9808017].
}

\lref\CarneirodaCunhaNW{
B.~G.~Carneiro da Cunha,
Phys.\ Rev.\ D {\bf 65}, 104025 (2002)
[arXiv:hep-th/0110169].
}

\lref\MaldacenaKR{
J.~M.~Maldacena,
arXiv:hep-th/0106112.
}

\lref\GiveonNS{
A.~Giveon, D.~Kutasov and N.~Seiberg,
Adv.\ Theor.\ Math.\ Phys.\  {\bf 2}, 733 (1998)
[arXiv:hep-th/9806194].
}

\lref\TeschnerFT{
J.~Teschner,
Nucl.\ Phys.\ B {\bf 546}, 390 (1999)
[arXiv:hep-th/9712256];
Nucl.\ Phys.\ B {\bf 571}, 555 (2000)
[arXiv:hep-th/9906215].
}

\lref\HorowitzJC{
G.~T.~Horowitz and D.~L.~Welch,
Phys.\ Rev.\ Lett.\  {\bf 71}, 328 (1993) [arXiv:hep-th/9302126];
%
N.~Kaloper,
Phys.\ Rev.\ D {\bf 48}, 2598 (1993) [arXiv:hep-th/9303007];
A.~Ali and A.~Kumar,
Mod.\ Phys.\ Lett.\ A {\bf 8}, 2045 (1993) [arXiv:hep-th/9303032];
%
M.~Natsuume and Y.~Satoh,
Int.\ J.\ Mod.\ Phys.\ A {\bf 13}, 1229 (1998)
[arXiv:hep-th/9611041];
Y.~Satoh,
Nucl.\ Phys.\ B {\bf 513}, 213 (1998) [arXiv:hep-th/9705208];
%
J.~M.~Maldacena and A.~Strominger,
JHEP {\bf 9812}, 005 (1998) [arXiv:hep-th/9804085];
%
S.~Hemming and E.~Keski-Vakkuri,
Nucl.\ Phys.\ B {\bf 626}, 363 (2002) [arXiv:hep-th/0110252];
%
 J.~Troost,
arXiv:hep-th/0206118;
%
E.~J.~Martinec and W.~McElgin,
JHEP {\bf 0204}, 029 (2002) [arXiv:hep-th/0106171];
}

\lref\BalasubramanianRY{
V.~Balasubramanian, S.~F.~Hassan, E.~Keski-Vakkuri and A.~Naqvi,
arXiv:hep-th/0202187.
}

\lref\CornalbaFI{
L.~Cornalba and M.~S.~Costa,
arXiv:hep-th/0203031.
}

\lref\NekrasovKF{
N.~A.~Nekrasov,
arXiv:hep-th/0203112.
}

\lref\SimonMA{
J.~Simon,
JHEP {\bf 0206}, 001 (2002)
[arXiv:hep-th/0203201].
}

\lref\LiuFT{
H.~Liu, G.~Moore and N.~Seiberg,
JHEP {\bf 0206}, 045 (2002)
[arXiv:hep-th/0204168];
arXiv:hep-th/0206182.
}

\lref\LawrenceAJ{
A.~Lawrence,
arXiv:hep-th/0205288.
}

\lref\FabingerKR{
M.~Fabinger and J.~McGreevy,
arXiv:hep-th/0206196.
}

\lref\HorowitzMW{
G.~T.~Horowitz and J.~Polchinski,
arXiv:hep-th/0206228.
}

\lref\SusskindQC{
L.~Susskind and J.~Uglum,
Nucl.\ Phys.\ Proc.\ Suppl.\  {\bf 45BC}, 115 (1996)
[arXiv:hep-th/9511227].
}

\lref\HartleAI{
J.~B.~Hartle and S.~W.~Hawking,
Phys.\ Rev.\ D {\bf 28}, 2960 (1983).
}

\lref\HartleTP{
J.~B.~Hartle and S.~W.~Hawking,
Phys.\ Rev.\ D {\bf 13}, 2188 (1976).
}

\lref\NiemiNF{
A.~J.~Niemi and G.~W.~Semenoff,
Annals Phys.\  {\bf 152}, 105 (1984).
}

\lref\HemmingKD{
S.~Hemming, E.~Keski-Vakkuri and P.~Kraus,
JHEP {\bf 0210}, 006 (2002)
[arXiv:hep-th/0208003].
}

\lref\DixonJW{
L.~J.~Dixon, J.~A.~Harvey, C.~Vafa and E.~Witten,
Nucl.\ Phys.\ B {\bf 261}, 678 (1985).
}

\lref\StromingerCZ{
A.~Strominger,
Nucl.\ Phys.\ B {\bf 451}, 96 (1995)
[arXiv:hep-th/9504090].
}

\lref\JohnsonQT{
C.~V.~Johnson, A.~W.~Peet and J.~Polchinski,
Phys.\ Rev.\ D {\bf 61}, 086001 (2000)
[arXiv:hep-th/9911161].
}

\lref\BanksVH{
T.~Banks, W.~Fischler, S.~H.~Shenker and L.~Susskind,
Phys.\ Rev.\ D {\bf 55}, 5112 (1997)
[arXiv:hep-th/9610043].
}

\lref\RastelliUV{
For a review of some recent developments see L.~Rastelli, A.~Sen and
B.~Zwiebach,
arXiv:hep-th/0106010.
}

\lref\WittenZW{
E.~Witten,
Adv.\ Theor.\ Math.\ Phys.\  {\bf 2}, 505 (1998)
[arXiv:hep-th/9803131].
}

\lref\BalasubramanianZV{
V.~Balasubramanian and S.~F.~Ross,
Phys.\ Rev.\ D {\bf 61}, 044007 (2000)
[arXiv:hep-th/9906226].
}
\lref\HubenyDG{
V.~E.~Hubeny,
arXiv:hep-th/0208047.
}

\lref\SusskindIF{
L.~Susskind, L.~Thorlacius and J.~Uglum,
Phys.\ Rev.\ D {\bf 48}, 3743 (1993)
[arXiv:hep-th/9306069].
}

\lref\GKS{N. Goheer, M. Kleban, and L. Susskind,
arXiv:hep-th/0212209.}

\lref\MaldacenaBW{
J.~M.~Maldacena and A.~Strominger,
JHEP {\bf 9812}, 005 (1998)
[arXiv:hep-th/9804085].
}

\lref\DysonNT{
L.~Dyson, J.~Lindesay and L.~Susskind,
JHEP {\bf 0208}, 045 (2002)
[arXiv:hep-th/0202163].
}

\newsec{Introduction}

It has been a long-standing goal of string/M theory to understand
the singularities in spacetime geometry that afflict classical
General Relativity. Much progress has been made in understanding
static, time independent,  singularities. For example, orbifolds
\DixonJW, conifolds \StromingerCZ,  and enhancons \JohnsonQT\ each
represent a successful resolution of a classical singularity, the
latter two requiring nonperturbative (in $g_s$) phenomena.

Much less is known about the fate of non-static, space-like or
null singularities. These are crucial in cosmology and include the
FRW big bang and big crunch singularities. The singularity at the
center of a black hole is of this type as well.  The conventional
wisdom has been that nonperturbative phenomena would come into
play near these singularities. Recently real calculations have
been done in perturbative string theory in mild big bang/big
crunch type backgrounds
\refs{\BalasubramanianRY,\CornalbaFI,\NekrasovKF,\SimonMA,
\LiuFT,\ElitzurRT,\CornalbaNV,\CrapsII,\FabingerKR }. In the
examples of cosmological singularities constructed  as time
dependent orbifolds of Minkowski space, the work of \refs{\LiuFT}
showed that tree level amplitudes diverge, due to infinite
blueshifts at the singularities. References
\refs{\LawrenceAJ,\HorowitzMW } discussed the physical meaning of
these results and argued that in general nonperturbative phenomena
should be expected around such points.

Enough progress has been made in string/M theory so that
algorithmically complete nonperturbative definitions of the theory
exist in certain backgrounds. These include Matrix Theory \BanksVH ,
the AdS/CFT correspondence \refs{\MaldacenaRE,\GubserBC,\WittenQJ}
 and its relatives, and, to some extent, String Field Theory
\RastelliUV. The
hope exists that such definitions could cast some light on the
space-like singularity problem.

The AdS/CFT correspondence seems particularly well suited to this
question because of the great success it has had in elucidating
the physics of black holes.  In particular the region outside the
horizon of an AdS- Schwarschild black hole is represented
holographically by the boundary CFT at finite temperature \WittenZW .

The black hole singularity is behind the horizon and so at first
glance the boundary CFT does not seem able to say anything about
it.   But on closer examination \refs{\BalasubramanianZV,
\MaldacenaKR,
\HubenyDG},
the boundary degrees of
freedom do seem to be able to probe the region of spacetime behind
the horizon, implementing the redundancy of description implied by
the ideas of black hole complementarity \SusskindIF. A particularly
clear example of this, building on an old observation of
Israel \IsraelUR,  involves the boundary description of an
eternal AdS-Schwarschild black hole.   Such a geometry has two
disconnected asymptotic boundaries, both approximately AdS.   Not
surprisingly, then, the holographic description of this geometry
involves {\it two} decoupled CFTs, one on each boundary
\refs{\HorowitzXK,\BalasubramanianDE,\MaldacenaKR}.  The only
coupling between CFTs is via the entangled state $|\Psi \rangle$,
referred to as the Hartle-Hawking state,  in which all expectation
values are taken. Correlation functions in one CFT reproduce the
thermal results for correlators outside the horizon of the black
hole; the black hole entropy in this formalism is the entanglement
entropy of the state $|\Psi \rangle$.\foot{This formalism has
recently been applied to the question of the quantum consistency
of de Sitter space by Goheer, Kleban, and Susskind \GKS.}
But correlation functions
involving the expectation values in $|\Psi \rangle$ of operators
in both CFTs should, as Maldacena \MaldacenaKR\ has argued,
contain some information about the geometry
behind the horizon.

The goal of this paper, building especially on the work of
\MaldacenaKR,
is to understand more carefully what kind
of behind the horizon information is contained in such correlators
and,
in
particular, what information about the singularity can be obtained
from them.

For simplicity we focus on the $2+1$ case \MaldacenaBW, $i.e.$, the
BTZ
black
hole \BanadosWN, which is an orbifold of AdS.  The spacelike
singularity of the nonextremal, nonrotating BTZ black hole is
given locally by a boost orbifold of two dimensional Minkowski
space times a spacelike line.   The two dimensional piece is
referred to as the Milne universe, and describes contracting and
expanding cones touching at a singularity. The null singularity
studied in \LiuFT\ is identical to the singularity of the zero
mass limit of the BTZ black hole.

The natural way to define Lorentzian correlators of boundary
operators in either the bulk or boundary description is via
analytic continuation from the euclidean theory.   Because of the
freedom to choose integration contours we show that it is possible
to describe a given amplitude as being determined by information
completely outside the horizon, or alternatively but equivalently
as being determined by information both inside and outside the
horizon.  This is reminiscent of the concept of black hole
complementarity \SusskindIF .\foot{Another indication of
complementarity in this
formalism has been discussed in \MaldacenaKR.}

As we will argue later, these continued amplitudes are expected to
be finite and the perturbation expansion for them well behaved.
We then must ask what happens to the breakdown expected
 from the singularities.
In the description involving data only outside the horizon there
is nothing to explain.  In the description that probes behind the
horizon we find, at least in one case, that the singular behavior
cancels between the future and past singularities.

Another question that arises concerns the intricate boundary
structure of Lorentzian BTZ.  We will argue that despite the
apparent existence of an infinite number of boundary components
the boundary CFTs only lie on the original two boundaries.

We now turn to a more detailed description of the content of the
paper.   Our main task is to show how to explicitly perform the
analytic continuation of  AdS/CFT amplitudes from Euclidean to
Lorentzian signature.  In principle, we could try to do this
directly at the level of the string worldsheet path integral, but
we will instead consider the simpler case of supergravity
amplitudes, as these are sufficient for our purposes.  The idea is
to start from some Euclidean supergravity amplitude, defined in
position space as an integration over the positions of interaction
vertices, which are in turn connected by various bulk-boundary and
bulk-bulk propagators.  The amplitudes are labelled by points on
the Euclidean boundary torus, corresponding to the locations of
operators in the boundary CFT.

As we proceed to continue the boundary points to the Lorentzian
section, we will have to deform the contour on which the
interaction vertices are integrated.  This is because the
propagators have singularities in the complex plane, and we must
deform the contour to avoid encountering the singularities.  We
obtain a Lorentzian interpretation from the form of the final
contour, as well as an \eps\ prescription for integrating around
the various singularities. Since there is some freedom in how we
deform the integration contour, there are a number of different
possible Lorentzian interpretations of the same analytically
continued amplitudes, of which we explore two.

The first corresponds to doing the natural contour deformation
with respect to integration over time in the BTZ coordinates,
which corresponds to a
Killing vector of the BTZ geometry.  This gives a Lorentzian
amplitude in which we integrate vertices over two coordinate
patches outside the horizon (the left and right wedges of the
Penrose diagram), as well as over two imaginary time segments
which can be thought of as imposing the Hartle-Hawking
wavefunction.   In this description, no explicit reference is made
to the region behind the horizon or to the singularity, and the
finiteness of the amplitudes is manifest. The analogous
continuation of boundary CFT amplitudes naturally leads to a
tensor product of two entangled CFTs associated to the boundaries
of the two coordinate patches, as in previous work
\refs{\HorowitzXK,\BalasubramanianDE,\MaldacenaKR,\HemmingKD }. So
the bulk and boundary description match up nicely, and in neither
do the other components of the BTZ geometry make an appearance.

The first continuation just described is analogous to continuing
flat space amplitudes with respect to Rindler time, whereas our
second continuation is analogous to continuing with respect to
Minkowski time.  For the latter case we introduce Kruskal
coordinates for BTZ, and perform the natural continuation with
respect to Kruskal time. This leads to a Lorentzian description in
which we integrate over a greater portion of the BTZ geometry than
before, including the BTZ singularity and beyond.  The \eps\
prescription provided by the analytic continuation tells us how to
integrate the vertices over the BTZ singularities.  Since we
effectively go around the singularity in the complex plane, a
naively  divergent result is replaced by a finite but complex
result. However, recalling that BTZ has both past and future
singularities, we show that unphysical imaginary parts cancel
betweeen the two singularities, at least in some cases.
The \eps\ prescription and the
cancellation between past and future singularities are the
mechanisms that seem to allow a well behaved boundary theory to
describe the singular geometry behind the horizon.

The possibility of choosing two different contours to describe the
same amplitude, one involving data only outside the horizon, the
other involving data behind the horizon, is reminiscent of black
hole complementarity ideas.  It is  striking that amplitudes
apparently related solely to phenomena outside the horizon can
also be used to reconstruct many properties of the geometry behind
the horizon as well as some other phenomena that occur there.
We note that these are phenomena that do not involve breakdown in the
perturbative description.

The remainder of this paper is organized as follows. In \S 2 we
review the BTZ geometry, its bulk-boundary propagator, and the
reason why we might expect divergences from the BTZ singularity.
In \S 3 we begin investigating the singularity with two point
functions and their relation to spacelike geodesics in the bulk,
although we later see that this approach has its limitations.
Arguments for the finiteness of analytically continued amplitudes
are given in \S 4. The prescription for the analytic continuation
from the point of view of boundary CFT is reviewed in \S 5.
Before discussing the analytic continuation in the bulk BTZ geometry,
sample computations in Minkowski spacetime
are given in \S 6.  Finally, in \S 7 we study the BTZ amplitudes
in two different ways by continuing with respect to BTZ time and
Kruskal time, and then discuss the results.

As this manuscript was being finished \Ben\ appeared, which has
significant
overlap with this work.

\newsec{Review of BTZ black hole}

\subsec{Geometry}

Let us recall the construction of the non-rotating BTZ black hole.
More details, including the rotating case, can be found in
\refs{\BanadosWN,\HemmingKD}.  Previous work on string theory on
BTZ includes
\refs{\HorowitzJC,\KeskiVakkuriNW,\MaldacenaKR,\HemmingKD}. The
starting point is \ads\ described as a hyperboloid embedded in a
flat spacetime with signature $(+,+,--)$:
\eqn\aa{x_0^2+x_1^2-x_2^2-x_3^2 =1.}
We are setting the \ads\ length scale to unity.  As usual, we will
actually work with the simply connected covering space of \aa. The
BTZ solution is obtained by identifying points by a boost,
\eqn\ab{  x_1 \pm x_2  ~\cong  ~e^{\pm 2\pi r_+} (x_1 \pm x_2).}
This will result in a non-rotating black hole of mass $M= r_+^2/8
G_N$. The line of fixed points at $x_1=x_2 =0$ is the black hole
singularity. The local geometry near the singularity is described
by the Milne universe times a line. Indeed, solving \aa\ for $x_3$
near the line of fixed points yields
\eqn\ac{ ds^2  \sim  -dx_1^2 +dx_2^2 +{dx_0^2 \over 1- x_0^2}.}
The boost identification in the $(x_1,x_2)$ plane  defines the
Milne universe.

To write coordinates that display the symmetries of the spacetime,
we break up \ads\ into the following three types of regions
\eqn\ad{\eqalign{{\rm Region~ 1:}\quad & x_1^2 - x_2^2 \geq 0,
\quad x_0^2 - x_3^2 \leq 0, \cr {\rm Region~ 2:}\quad & x_1^2 -
x_2^2 \geq 0, \quad x_0^2 - x_3^2 \geq 0, \cr {\rm Region~
3:}\quad & x_1^2 - x_2^2 \leq 0, \quad x_0^2 - x_3^2 \geq 0.}}
We then cover each region by four separate coordinate patches,
corresponding to the values of $\eta_{1,2} =\pm 1$,

\noindent {\bf Region 1:}
\eqn\ae{\eqalign{ x_1 \pm x_2  & = \eta_1  {r \over r_+} e^{\pm
r_+ \phi} \cr x_3 \pm x_0 & = \eta_2 {\sqrt{r^2 - r_+^2} \over r_+
}e^{\pm r_+ t} . }}

\noindent {\bf Region 2:}
\eqn\af{\eqalign{ x_1 \pm x_2  & = \eta_1  {r \over r_+} e^{\pm
r_+ \phi} \cr x_3 \pm x_0 & = \eta_2 {\sqrt{r_+^2 - r^2} \over r_+
}e^{\pm r_+ t} . }}

\noindent {\bf Region 3:}
\eqn\ag{\eqalign{ x_1 \pm x_2  & = \eta_1 {\sqrt{r^2 - r_+^2}
\over r_+ }e^{\pm r_+ t}  \cr x_3 \pm x_0 & = \eta_2 {r \over r_+}
e^{\pm r_+ \phi} . }}

$r$ lives in the range $(r_+,\infty)$ in regions 1 and 3, and
$(0,r_+)$ in region 2.  The BTZ identification in these
coordinates is
\eqn\ah{\eqalign{{\rm Regions ~1,2:}\quad & (t,\phi,r) ~ \cong ~
(t,\phi+2\pi,r) \cr {\rm Region ~3:}\quad & (t,\phi,r) ~ \cong ~
(t+2\pi,\phi,r). }}
The metric is
\eqn\at{ds^2 = - (r^2-r_+^2) dt^2 + { dr^2 \over r^2-r_+^2} + r^2
d\phi^2.}
In string theory there is also a nonvanishing B-field, but we will
not need its explicit form.

Noting that $t$ is a timelike coordinate in region 3, we see that
the BTZ identification \ah\ gives rise to closed timelike curves
in this region.  The desire to avoid these motivated the proposal
to truncate the geometry at the singularity \BanadosWN.  One goal
of the present work is to examine whether such a truncation
actually occurs in the context of string theory and the AdS/CFT
correspondence.

To get a picture of  the global structure, it is helpful to display
two orthogonal cross sections of the original \ads\ cylinder in Fig.
1,
with
the various coordinate regions indicated.
Important for us is the fact that each component of regions 1 and
3 has a distinct boundary.  One might then expect there to be
distinct CFT's living on each boundary component;  we will see in
\S 7 that the actual situation is more subtle.

\fig{Two orthogonal cross sections of the \ads\ cylinder, with BTZ
coordinate patches indicated.  Both diagrams should be extended
periodically in
the vertical direction.
}{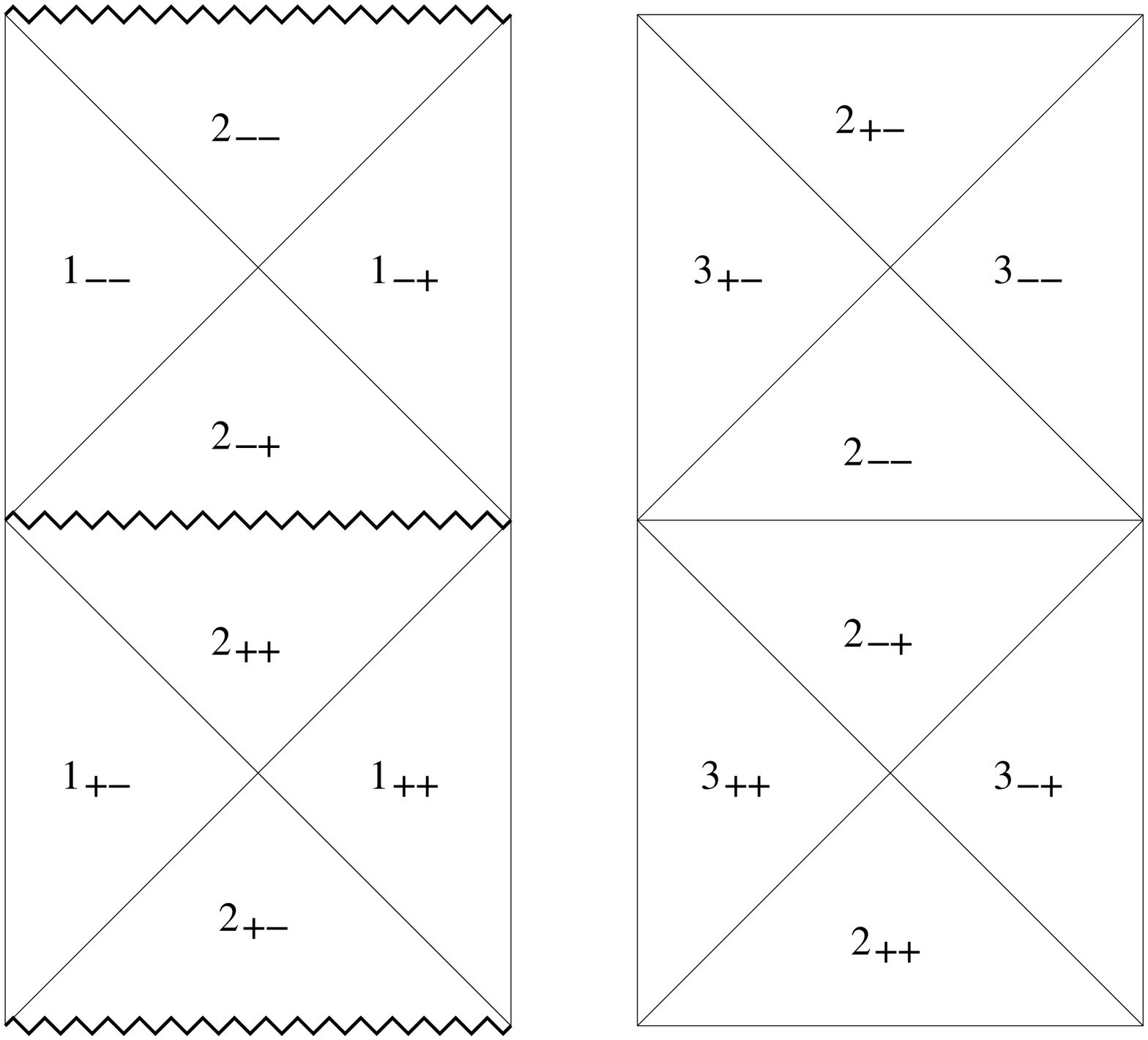}{4.5truein}

\subsec{Propagators and divergences}

AdS/CFT correlation functions on the bulk side are constructed out
of bulk-boundary and bulk-bulk propagators.  The BTZ versions of
these propagators can be obtained from their \ads\ cousins by the
method of images \KeskiVakkuriNW. For the bulk-boundary propagator
we need to specify a ``source'' point $b$ on the boundary, and a
``sink'' point $x$ in the bulk. In BTZ coordinates, the form of
the propagator changes as we move the source and sink points from
one region to another. For a minimally coupled scalar of mass $m$,
the bulk-boundary propagator for both source and sink in region
$1_{++}$ is, up to normalization,
\eqn\ai{K^{(1_{++}1_{++})}(x,b') = \sum_{n=-\infty}^\infty {1\over
\left[-\sqrt{{r^2-r_+^2 \over r_+^2}}\cosh(r_+ \Delta t)+{r\over
r_+}\cosh r_+(\Delta \phi+2\pi n)\right]^{2h_+}}~. }
Here $\Delta t = t-t'$, and similarly for $\Delta \phi$.  A bulk
scalar of mass $m$ corresponds to a boundary operator of conformal
dimension $2h_+ = 1 + \sqrt{1+m^2}$.  Although written for region
$1_{++}$, in fact $K$ is always given by \ai\ whenever the source
and sink point are in the same region. \ai\ diverges when the bulk
and boundary points are lightlike separated, and so an $i\epsilon$
prescription
is required.   We will see how to obtain the correct \eps\
prescription when we discuss the continuation from Euclidean
signature.

To move the sink point to another region, we can analytically
continue the propagator.  For instance, by examining \ae, we see
that to move the sink point to $1_{+-}$ we should make the
replacement $t \rightarrow t - i\pi/r_+$.   Note that the
imaginary shift is half the inverse Hawking temperature,
\eqn\aii{ \beta = 1/T_H = 2\pi /r_+ .}
 Making this replacement, the
bulk-boundary propagator becomes
\eqn\ak{K^{(1_{+-}1_{++})}(x,b') = \sum_{n=-\infty}^\infty {1\over
\left[\sqrt{{r^2-r_+^2 \over r_+^2}}\cosh(r_+ \Delta t)+{r\over
r_+}\cosh r_+(\Delta \phi+2\pi n)\right]^{2h_+}} . }
This propagator is nonsingular, reflecting the fact that $1_{++}$ and
$1_{+-}$ are spacelike separated.

To investigate the behavior of the propagator near the BTZ singularity
we now move the sink point to $2_{++}$,
\eqn\al{K^{(2_{++}1_{++})}(x,b') = \sum_{n=-\infty}^\infty {1\over
\left[-\sqrt{{r_+^2-r^2 \over r_+^2}}\sinh(r_+ \Delta t)+{r\over
r_+}\cosh r_+(\Delta \phi+2\pi n)\right]^{2h_+}} . }
This is singular at $r=0$, since the summation then acts on an $n$
independent quantity.   By estimating the number of term in the sum
which contribute as $r \rightarrow 0$, we find that near the
singularity
\eqn\am{ K^{(2_{++}1_{++})} ~\sim ~ f(\Delta t) \ln r.}
The same divergence applies when we approach the singularity from
other
regions.  The bulk-bulk propagators also diverge logarithmically for
the same reason.

Let us first be very naive and see why we might expect divergent
amplitudes to arise.
A typical supergravity calculation of an AdS/CFT amplitude
involves a Feynman diagram composed  of propagators and vertices, and
an
integration over the positions of the vertices.  Divergences can
therefore
arise from the region of integration involving some number of vertices
approaching the BTZ singularity.  In fact, since the integration
measure
is  $\int \! dt \, d\phi \, dr \, r$,
nonderivative couplings will yield finite amplitudes after
integration.
However, an interaction with a sufficient number of derivatives will
lead to a divergent integral.  The divergences arise due to an
infinite
blueshifting at the singularity, as in recent examples of time
dependent
orbifolds of Minkowski space. For the Milne singularity, divergences
in string amplitudes are studied in \Ben. In \S 7 we will see how the
AdS/CFT correspondence handles these divergences.

\newsec{Probing the singularity with spacelike geodesics}

{} From our knowledge of the bulk-boundary propagator in the various
regions, we can make a few preliminary comments about how AdS/CFT
correlators might probe the singularity. We will see later that the
situation is considerably more subtle than these considerations
suggest.  As a specific example, consider a two point function with one
operator inserted on the boundary of $1_{++}$ and another on the
boundary of $1_{+-}$.  According to the standard AdS/CFT rules,
\ak\ leads to the two point function \MaldacenaKR
\eqn\an{\langle {\cal O}_{1_{+-}} {\cal O}_{1_{++}} \rangle
~=~\sum_{n=-\infty}^\infty \left[ \cosh (r_+ \Delta t) +
\cosh r_+ (\Delta \phi + 2\pi n) \right]^{-2h_+}.}

Given the BTZ causal structure, correlators involving operators in
both $1_{++}$ and $1_{+-}$ might be expected to probe physics
behind the horizon and in particular near the singularity.  We can
make this expectation a bit more precise by using the WKB
approximation to see which spacetime geodesics contribute to \an.
Consider for simplicity the two-point function with $\Delta \phi
=0$.  The equation for a spacelike geodesic is
\eqn\ao{ {\dot{r}^2 \over r^2 -r_+^2} -{E^2 \over r^2-r_+^2} =1,}
where $\dot{}$ denotes a proper time derivative and $E$ is the
conserved energy, $  E = (r^2 - r_{+}^{2}) \dot{t}$. Integrating
we find
\eqn\ec{ r(\tau) =\left\{ \eqalign{\pm &
\sqrt{E^2-r_+^2}\sinh(\tau-\tau_0), \quad\quad  E^2 > r_+^2 \cr
\pm & \sqrt{r_+^2-E^2}\cosh(\tau-\tau_0), \quad\quad  E^2 < r_+^2.
}\right.}
For $E^2 > r_+^2$ the geodesics cross the singularity at $r=0$,
and so we  focus on the $E^2 < r_+^2$ case and choose the $+$
sign.
 The distance
of closest approach to the singularity is
\eqn\ee{ r_{min} = \sqrt{r_+^2 - E^2}.}
We want to relate $E$ to the values of the boundary time
coordinates.  Integrating the equations for $t$ gives \eqn\eg{
\Delta t =t(\infty) - t(-\infty) = -{i\pi \over r_+} +{1 \over
r_+} \ln \left\{ {1+{E \over r_+} \over 1 - {E\over r_+}}
\right\}.}
The imaginary part, $-i\beta/2$, is the correct jump when going
between $1_{++}$ and $1_{+-}$. The real part, $\Delta t_r$, is
related to $r_{min}$ by
\eqn\ei{r_{min} = {r_+ \over \cosh\left({r_+ \Delta t_r \over 2}
\right)}.}
In our conventions time runs backward in $1_{+-}$, which is
consistent with the fact that $r_{min}$ is invariant under
simultaneous time translations in the initial and final times.

The WKB approximation to the two-point function is given by
$e^{-S}$, where $S$ is the action of the spacelike geodesic
passing between the two boundary points.   This action is
divergent; using a large $r$ cutoff the regularized action is
\eqn\fb{ S= m\Delta \tau = 2m\ln \left({ 2 r_c \cosh\left({r_+
\Delta t_r \over 2}\right) \over r_+} \right).}
We define a renormalized action $S_{{\rm ren}}$ by subtracting $2m
\ln r_c$, since this term arises also in pure \ads. The WKB
approximation to the two point function  is then
\eqn\fc{e^{-S_{{\rm ren}}} = {C \over \left(\cosh\left({ r_+
\Delta t_r \over 2}\right)\right)^{2m}}.}
When we recall that for large $m$, which is when  the WKB
approximation is accurate,  $2h_+ = 1+\sqrt{1+m^2} \approx m$, we
find that \fc\ agrees with the leading term in \an.

We can ask for the time scale at which the geodesic passes within
a proper distance $L_{{\rm Pl}}$ of the singularity, which is when
we could expect quantum effects to become important. Restoring the
\ads\ length scale,  this is
\eqn\ej{ \Delta t_{{\rm sing}} ~\sim ~{2 L_{{\rm AdS}}^2 \over
r_+} \ln\left( {L_{{\rm AdS}} \over L_{{\rm Pl}}} \right).}

Another important timescale was pointed out by Maldacena
\MaldacenaKR.  This is the timescale where large fluctuations in the
geometry apparently become important.  For sufficiently large time
separation, $\Delta t > \Delta t_{{\rm fluc}}$,  \fc\ is
inconsistent with unitarity of the boundary theory, since the
correlation should not drop below $e^{-s}$, where $s$ is the
entropy,
\eqn\ap{s = {\pi r_+ \over 2   L_{{\rm Pl}}}.}
This gives
\eqn\eq{ \Delta t_{{\rm fluc}} ~\sim~ {\pi \over 2}{L_{{\rm AdS}}
\over  m L_{{\rm Pl}}}.}

The time $\Delta t_{{\rm fluc}}$ marks the onset of fluctuations in
the
correlation function
of size $\sim \exp(-s)$.  A much longer time, the Poincar\'e
recurrence
time,
$\Delta t_{{\rm recur}} \sim \exp(a s)$, marks the onset of order one
fluctuations
\refs{\MaldacenaKR, \DysonNT}.
{}From the bulk point of view, an indication of the timescale for
fluctuations can be
seen in the WKB approximation when we recall that we should really
consider the sum of the actions of the geodesic and the background
geometry.  The action of the black hole is related to its free
energy
\eqn\ar{ S_{{\rm bh}} = -(s - \beta M) = -{\pi \over 4} { r_+\over
L_{{\rm Pl}}} .}
On the other hand, recalling that pure \ads\ has energy $M =
-1/8L_{{\rm Pl}}$, the action of thermal \ads\ at inverse
temperature $\beta$ is
\eqn\as{S_{{\rm AdS}} = \beta M = - {\pi \over 4}{L_{{\rm AdS}}^2
\over  L_{{\rm Pl}} r_+ }.}
So for $r_+ > L_{{\rm AdS}}$ the black hole dominates the
partition sum.  However, this can be overcome by the positive
action of the action for the spacelike geodesic.  Indeed, the
timescale for the geodesic action  to become comparable to the
black hole action recovers (up to a numerical factor) the result
\eq.

Comparing \ej\ with \eq, we see that for $m \sim L_{{\rm AdS}}$,
$r_+ \sim L_{{\rm AdS}} \gg L_{{\rm Pl}}$,  we have $\Delta
t_{{\rm fluc}} \gg \Delta t_{{\rm sing}}$.  Therefore, we might
hope to use boundary correlators to probe the physics of the
singularity before possible fluctuations in the whole geometry become
important. This will turn
out to be only indirectly the case.

\newsec{Analytic continuation I: finiteness of amplitudes}

The heuristic arguments just given are not sufficient to determine
to what extent we can really probe the singularity.  The
divergences arising in time dependent orbifolds of Minkowksi space
have to do with interactions near the singularity.  Similarly, in
the BTZ case we need to go beyond the two-point function and
include interactions in the bulk.

At our current level of understanding, string theory in Lorentzian
\ads\ or BTZ is defined by analytic continuation from Euclidean
signature \refs{\GiveonNS,\MaldacenaHW,\TeschnerFT}. This is the
approach we will follow; we will discuss later whether this
procedure really captures all of the Lorentzian physics.

The Euclidean BTZ metric is given by the replacement $t= -i\tau$,
\eqn\ba{ ds^2 = (r^2-r_+^2) d\tau^2 + {dr^2 \over r^2- r_+^2} +r^2
d\phi^2,}
with
\eqn\bb{\tau \cong \tau + \beta}
and $\beta$ given by \aii. The radial coordinate is now restricted
to $(r_+,\infty)$. Given the periodicity of $\tau$ and $\phi$,
\ba\ is topologically a solid torus.   The boundary CFT therefore
lives on a torus parameterized by $\tau$ and $\phi$.

A Euclidean AdS/CFT $n$-point function is labelled by $n$ points
on the boundary torus,  $G_n(\tau_1,\phi_1, \ldots , \tau_n,
\phi_n)$. The amplitudes are initially defined for real $\tau$, or
equivalently for pure imaginary $t$.  To obtain Lorentzian
amplitudes we need to perform an analytic continuation in $t$.
Continuing a point to real $t$ gives a point on a boundary
component of Lorentzian BTZ, which we can take to be in $1_{++}$.
As we have already mentioned, to get from $1_{++}$ to $1_{+-}$ one
takes $t \rightarrow t - i\beta /2$.  So, starting from $t$ on the
imaginary axis we need the continuations
\eqn\bc{ t ~ \rightarrow ~ \left\{ \eqalign{ &{\rm real}
\quad\quad ~~~~~~~~~ 1_{++} \cr  & {\rm real} - i\beta/2
\quad\quad 1_{+-} } \right. }
We will defer to later the question of continuing to other
boundary components.

We now want to argue that the analytically continued amplitudes
are finite.  The argument can be made in terms of either the bulk
or boundary descriptions. From the boundary point of view, since
we know that our amplitudes correspond to those of a well behaved
CFT on the boundary torus, we do not expect there to arise any
unusual singularities in amplitudes
even after analytic continuation. We expect correlation functions
defined for real $t$ to be analytic in $t$, order by order in the
string loop counting parameter.   This follows from a well behaved
spectral decomposition (a natural expectation) or from the
perturbative bulk correspondence. This analyticity implies that
singularities will be at most complex codimension one. But the
kind of singularities induced by effects like \am\ will in general
be of real codimension one.\foot{The tree level LMS amplitudes
\LiuFT\ have singularities only at complex codimension one, but
higher orders are expected to be generically singular
\refs{\LawrenceAJ,\HorowitzMW}.} Another way of saying this is
that, as we review in the next section,  Lorentzian amplitudes are
manifestly regular and finite since they can be expressed as
expectation values evaluated in the entangled state,
\eqn\bca{  |\Psi \rangle = {1\over \sqrt{{\cal Z}}} \sum_n
e^{-\beta E_n/2} |n\rangle \otimes
               |n \rangle,}
where $|n \rangle$  is an energy eigenstate with energy $E_n$ in
the Hilbert space of the CFT and ${\cal{Z}}$ is the partition
function.

{}From the bulk point of view, the basic point is that the Euclidean
BTZ geometry is completely smooth, as usual for Euclidean black
holes, since the region $r < r_+$ does not appear. Therefore,
string theory or supergravity amplitudes computed in Euclidean
signature will be finite, modulo the usual divergences that  occur
even for pure \ads, such as due to tachyons and so forth, and can
be analytically continued to Lorentzian signature as above.  One
may think that there is a possibility that these amplitudes do not
have good asymptotic expansions in the string coupling constant.
This, however, is not likely. Since the BTZ geometry is an
orbifold of $AdS_3$, at the tree level, a correlation function in
the former can be expressed as a sum over the corresponding
correlation function in the latter under the action of the
orbifold group. This sum is manifestly convergent \MaldacenaKR.
Moreover, as we will see in the next section, correlation
functions of operators on $1_{++}$ and $1_{+-}$ can be computed
taking into account interactions taking place outside of the
horizon only. Thus we do not expect divergences associated to the
singularity to arise at higher loops either.   Of course field
theoretic
divergences could be rendered finite by stringy $\alpha'$ effects, but
this seems
unlikely, especially given the stringy divergences found in \LiuFT .

\newsec{Analytic continuation II: boundary theory}

Analytic continuation from Euclidean signature yields finite
amplitudes, and we now want to examine in more detail how this
comes about.  As we discussed previously, Lorentzian signature
divergences seemingly arise from integrating an interaction vertex
near the BTZ singularity.  We will find two different
interpretations, corresponding to two different contour
deformations, for how the singularity is avoided.  In the first,
interactions only occur in regions $1_{++}$ and $1_{+-}$, so that
the region near the singularity never appears in the calculation.
In the second interpretation the region near the singularity does
appear, but the analytic continuation provides an \eps\
prescription which tells us how to go around the singularity in
the complex plane.

It is useful to begin by reviewing the analytic continuation in
the boundary theory, following the work of Niemi and Semenoff
\NiemiNF. For simplicity, we consider a weakly interacting scalar
field theory on the Euclidean torus.  We consider the computation
of Euclidean time ordered correlation functions
\eqn\ca{\eqalign{G_n(\tau_1,\phi_1, \ldots , \tau_n, \phi_n) & =
{\rm Tr}\left\{ e^{-\beta H} {\rm T}\left[X(\tau_1,\phi_1), \ldots
, X(\tau_n, \phi_n)\right]\right\} \cr  & \cr&= \int_{\rm
periodic} {\cal D}X \, e^{-S} ~X(\tau_1,\phi_1), \ldots ,
X(\tau_n, \phi_n).}}
We imagine computing Feynman diagrams in position space, so we
will have interaction vertices integrated over the Euclidean
torus.  A simple example is the lowest order three-point function
in the presence of a $\lambda X^3$ interaction,

\eqn\cb{ G_3(\tau_1,\phi_1,\tau_2,\phi_2,\tau_3,\phi_3) ~\sim ~
\lambda \int_0^\beta \! d\tau \int_0^{2\pi} \! d\phi \,
G(\tau,\phi,\tau_1,\phi_1)G(\tau,\phi,\tau_2,\phi_2)G(\tau,\phi,\tau_3,\phi_3).}

Now relabel $\tau_i = it_i$ and $\tau = it$ and consider
analytically continuing $G_n$ to the real $t_i$ axis.  The point
is that the propagators have singularities for lightlike separated
arguments.  The positions of these singularities in the complex
$t$ plane will move around as we continue in $t_i$, and we have to
deform the contour of integration so that no singularities cross
the contour.   Singularities occur for
\eqn\cc{ t = t_i \pm (\phi - \phi_i) + in\beta, \quad\quad n= {\rm
integer}.}
The $t$ contour of integration originally runs from $0$ to
$-i\beta$ along the imaginary axis.  It is convenient to use
translation invariance to instead take the contour to run from
$-T$ to $-T -i\beta$ with $T$ real and positive.  Eventually, we
will take $T \rightarrow \infty$.

So before doing any analytic continuation,  Fig. (2a)  shows
the integration contour and the locations of singularities in the
integrand.

\fig{Integration contours for evaluating correlation functions.
Contour
(a) defines a Euclidean amplitude; analytic continuation to real
time gives (b).
}{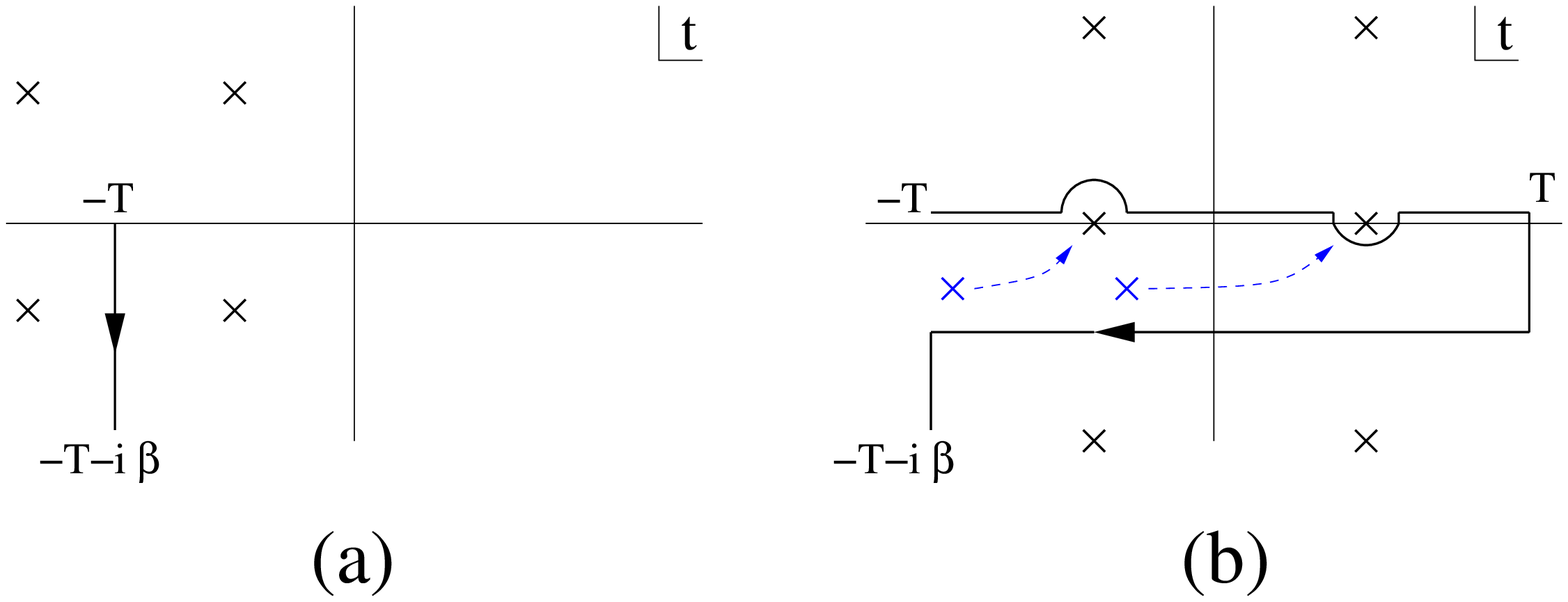}{5.5truein}

We have only drawn the singularities due to a single propagator to
avoid clutter.  Now move $t_i$ to the real axis.   The locations
of singularities move according to \cc.  Deforming the contour to
avoid the singularities, we end up with the contour in Fig. (2b).
We are left with two segments parallel to the real axis, as well
as two segments parallel to the imaginary axis.  Singularities on
the real axis are avoided by the usual prescription leading to the
Feynman Green's function.

The result has a simple operator interpretation.  The two
horizontal segments represent two Lorentzian copies of the
original Euclidean field theory. The Hilbert space of the theory
is ${\cal H} \otimes {\cal H}$ where ${\cal H}$ is the Hilbert
space of the field theory on the cylinder.  The two copies
communicate via the vertical segments. The vertical segments
represent insertions of the operator $e^{-\beta H/2}$,
corresponding to an imaginary time translation by $\beta/2$.

More precisely, the result can be written in operator form as
\eqn\cd{G_n = \langle \Psi |{\rm T}\left[X(t_1,\phi_1), \ldots ,
X(t_n, \phi_n)\right]|\Psi \rangle,}
where $|\Psi \rangle$ is an entangled state in ${\cal H} \otimes
{\cal H}$,
\eqn\ce{ |\Psi \rangle = {1\over \sqrt{{\cal Z}}} \sum_n e^{-\beta
E_n/2} |n\rangle \otimes
               |n \rangle.}
$T$ in \cd\ now represents Lorentzian time ordering.  Since we
have continued to the real $t$ axis, the $X$ operators in \cd\ all
represent operators in a single copy of the field theory, say the
first.  It is clear that we can then perform the trace over states
in the second copy, and recover a thermal expectation value for
operators in the first copy,
\eqn\cf{G_n = {\rm Tr}\left\{ e^{-\beta H} {\rm
T}\left[X(t_1,\phi_1), \ldots , X(t_n, \phi_n)\right] \right\}.}

It is straightforward to generalize the previous argument to the
case where some operators are continued to
$t = {\rm real} -i\beta
/2$.   The resulting contour appears as in Fig. 3.
\fig{Time integration contour for operators on both Lorentzian copies.
}{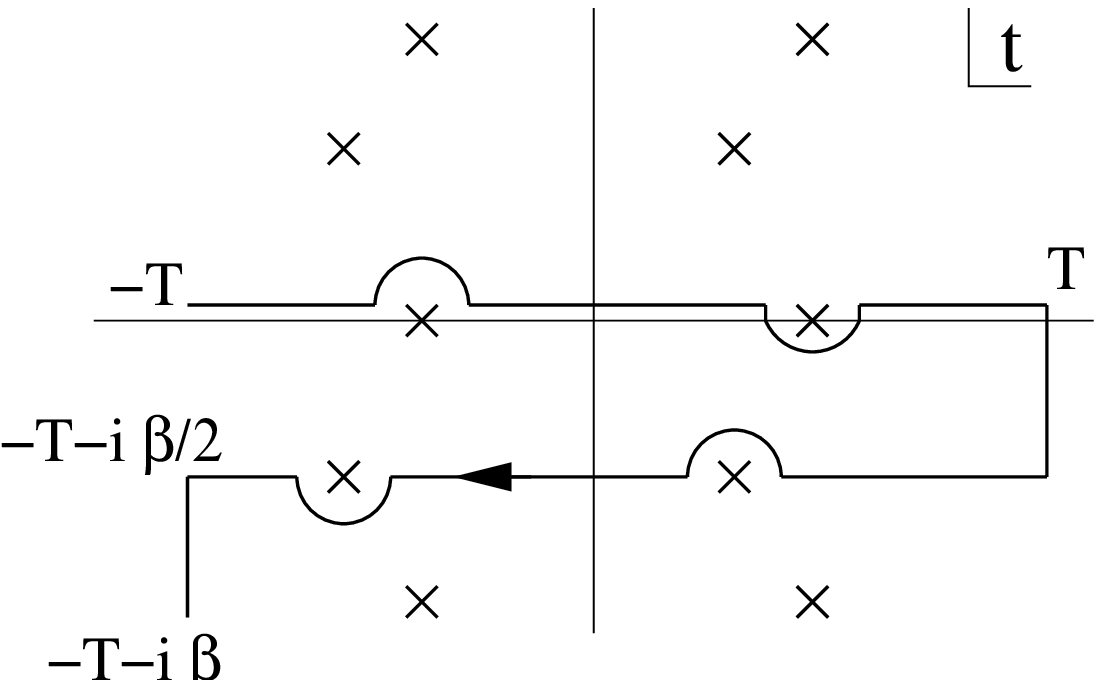}{2.75truein}

The expression \cd\ is unchanged, except that now whichever
operators  were taken to $t= {\rm real} -i\beta/2$ now appear as
operators in the second copy of the field theory. Finally, we can
also consider continuing operators to the vertical segments of the
final contour.  This has the effect of replacing $e^{-\beta H/2}$
by a more general operator, and so corresponds to changing the
state from \ce\ to something else.

Let us make a few comments about these results.  First, although
we only explicitly discussed the continuation of diagrams with a
single vertex, the argument is easily generalized by considering
each vertex in turn. Second, it is important to note that the
continuation instructs us to integrate vertices over the entire
contour, including the vertical segments.  The presence of
interactions on the vertical segments ensures that the energy
eigenstates appearing in  \ce\ are the correct energy eigenstates
of the full interacting theory.  Integrating only over the
horizontal segments would yield energy eigenstates of the free
theory.

As noted by Israel \IsraelUR\ shortly after Hawking's derivation
of black hole radiance (and  in the context of AdS/CFT in
\refs{\HorowitzXK,\BalasubramanianDE,\MaldacenaKR}), the fact that
real time thermal correlators are naturally interpreted in terms
of a tensor product of two field theories is directly analogous to
the fact that constant time hypersurfaces in an eternal  black
hole geometry naturally consist of two components on either side
of the horizon.   In our notation, the two components correspond
to $1_{++}$ and $1_{+-}$. So the expectation that there should be
two boundary theories associated with the two boundaries of
$1_{++}$ and $1_{+-}$ is borne out by analytic continuation.

\newsec{Analytic continuation III: Minkowski space}

There is some additional freedom to analytically continue bulk
amplitudes corresponding to different choices of time coordinates.
Different choices will lead to different Lorentzian
interpretations of the same correlation functions.  Before
proceeding to the black hole case we will do a  warmup
example.

We start by computing Green's functions in flat Euclidean space
\eqn\da{ ds^2 = d\tau^2 + dx^2.}
So, for example,  the expression analogous to \cb\ is now
\eqn\db{ G_3(\tau_1,x_1,\tau_2,x_2,\tau_3,x_3) ~\sim ~ \lambda
\int_{-\infty}^\infty \! d\tau \int_{-\infty}^\infty \! dx \,
G(\tau,x,\tau_1,x_1)G(\tau,x,\tau_2,x_2)G(\tau,x,\tau_3,x_3).}
The standard procedure is to continue in $t_i=-i\tau_i$ while
rotating the time contour to the real $t$ axis.  An \eps\
prescription follows from taking the contour to be at a small
angle with respect to the real axis, or equivalently, to go around the
singularities as in Fig. 4.

\fig{Standard contour rotation defining amplitudes in Minkowski space.
}{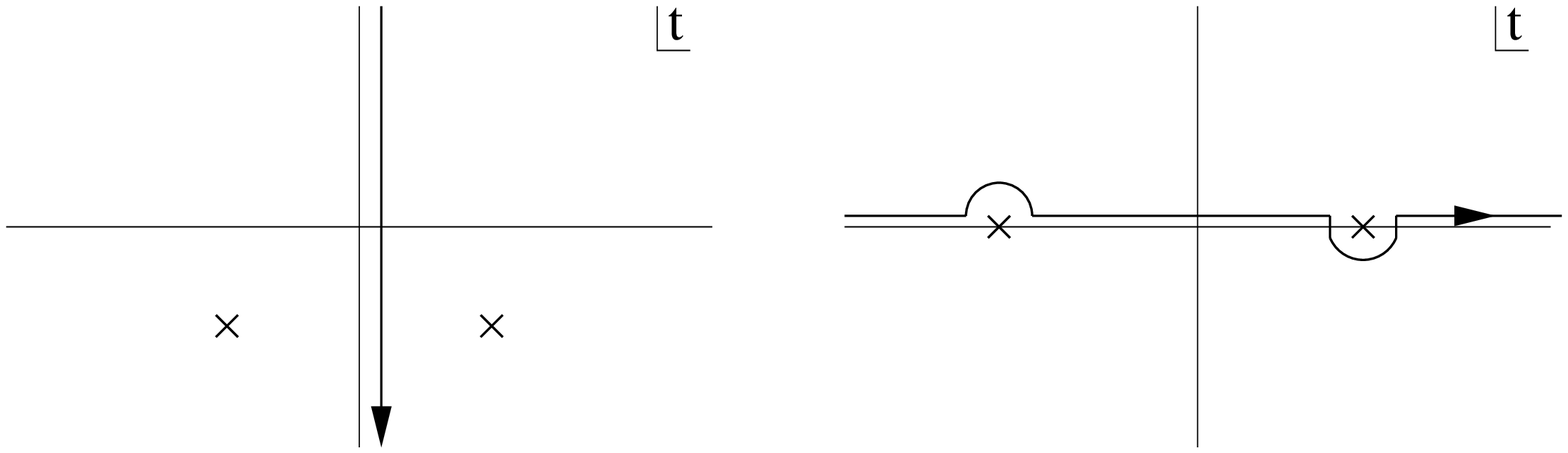}{4.5truein}
The result is that we are to integrate vertices over Minkowski
spacetime using the Lorentzian propagator
\eqn\dc{G_{{\rm Lor}}(t,x,t',\phi') =  G(e^{i(\pi/2
-\epsilon)}t,x,e^{i(\pi/2 -\epsilon)}t',\phi').}
Since the Euclidean propagator is a function of $\sigma^2 =
(\tau-\tau')^2 + (x-x')^2$, the rule to obtain the Lorentzian
propagator is
\eqn\dd{\sigma^2 \rightarrow -(t-t')^2 + (x-x')^2 + i\epsilon.}

We can alternatively analytically continue with respect to Rindler
time.  To do this we transform to polar coordinates
\eqn\de{ \tau = r\sin \theta, \quad x= r \cos \theta, \quad ds^2 =
dr^2 + r^2 d\theta^2.}
The Euclidean integration is now $\int_0^\infty dr~r
\int_{0}^{2\pi} d\theta$.

Recall that Rindler coordinates cover Minkowski spacetime in four
patches,
\eqn\df{ x \pm t = \left\{ \eqalign{ & ~~~~r e^{\pm
\eta}\quad\quad R \cr & -r e^{\pm \eta}\quad\quad L \cr & ~~~~r
e^{\mp \eta}\quad\quad F \cr & -r e^{\mp  \eta}\quad\quad P
}\right. }
with metric
\eqn\dg{ ds^2 = \left\{ \eqalign{ &-r^2 d\eta^2 + dr^2 \quad\quad
R,L \cr & ~~~~r^2 d\eta^2 -dr^2 \quad\quad F,P } \right.}
Note that region $L$ is obtained from region $R$ by $\eta
\rightarrow \eta - i\pi$.   We will take $\eta = -i\theta$  to be
the Rindler coordinate in region R.
\fig{Rindler coordinate patches.
}{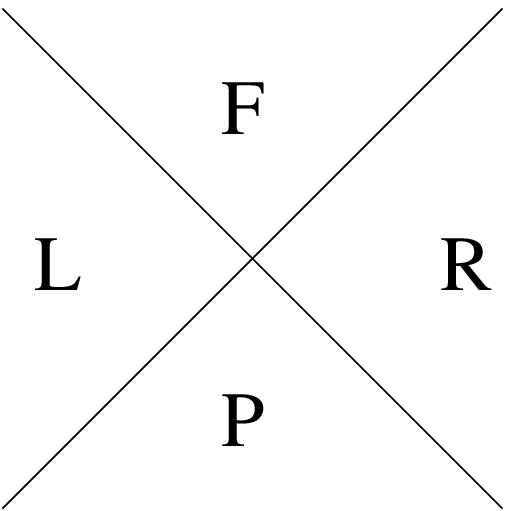}{1.0truein}

Now, the geodesic distance expressed in terms of $r$ and $\eta$ is
\eqn\dh{ \sigma^2 = r^2 +r'^2 -2 r r' \cosh(\eta-\eta').}
Therefore, singularities in the complex $\eta$ plane are located
at
\eqn\di{\eta = \eta_i +{ 1 \over 2 r r_i} \cosh^{-1}(r^2 + r_i^2)
= \eta _i  + 2\pi i n +\pm {\rm real}.}
With $\beta =2\pi$, our integration contour in the $\eta$ plane
and the location of singularities are precisely the same as in our
earlier discussion of continuing correlators on the Euclidean
cylinder.  Therefore, we can deform the contour as in Fig. 2 (with
$t$ replaced by $\eta$).
  The two horizontal segments now correspond to
integration over region $R$ and $L$.  The appearance of a tensor
product is now seen to be due to the fact that the $t=0$ Minkowski
timeslice is a sum of $\eta=0$ timeslices in the right and left
Rindler patches.

Green's function computed by continuation in either Minkowski or
Rindler time should agree, and this indeed follows from the fact
that the entangled state arising in the Rindler description
\eqn\dj{|\Psi \rangle = {1 \over \sqrt{{\cal Z}}} \sum_n e^{-\pi
E_n} |n\rangle_R  \otimes |n\rangle_L  }
is equal to the usual Minkowski vacuum \UnruhDB.  To see that the two
states are the same, consider a path integral on the lower half
Euclidean plane, with prescribed boundary conditions $\phi(x)$
on the real axis.
This wavefunctional $\Psi[\phi(x)]$
defines the Minkowski vacuum state.  On the other
hand, we can consider the Hilbert space of wavefunctions on half of
the
real axis, with a Rindler Hamiltonian $H_R$ corresponding to rotations
about
the real axis.   The path integral then becomes the transition
amplitude
$\langle \phi_L | e^{-\pi H_R} | \phi_R \rangle$, where $\phi_{L,R}$
are
the boundary conditions $\phi(x)$ restricted to the left and right
halves
of the real axis.  Inserting a complete set of eigenstates of
 $H_R$ then leads to the equivalence of the two states.

So to summarize, Green's functions with arguments in regions $R$
and $L$ can be computed either in the usual fashion by integrating
vertices over all of Minkowski space, or by just integrating over
the $R$ and $L$ wedges with an entanglement given by \dj.  If we
imagine first doing the integration over the vertical segments of
the Rindler contour, this will result in wavefunctions inserted at
$\eta = \pm \infty$.   These wavefunctions provide the boundary
conditions at the horizons which bound the two Rindler wedges.
Equivalently, the wavefunctions can be thought of as providing the
``missing'' part of the integrand from not integrating over the
$F$ and $P$ wedges.

\newsec{Analytic continuation IV: behind the black hole horizon }

Now we are ready to discuss analytic continuation to compute
correlation functions in the Lorentzian BTZ black hole.

\subsec{BTZ coordinates}

We first consider analytic continuation in BTZ coordinates \at.
This is straightforward and follows closely our discussion of
analytic continuation in Rindler time.  Singularities in propagators,
occuring, as always, for lightlike separation, are located in the
complex time plane at
\eqn\ha{ t = t' +  i m \beta \pm {\rm real}.}
For instance, for the bulk-boundary propagator given in \ai\ the
singularities
are located at
\eqn\hb{ t =t' + i m \beta + \cosh^{-1} \left(
\sqrt{{r^2 \over r^2 -r_+^2}}\cosh r_+(\Delta \phi + 2\pi n) \right).}
Euclidean AdS/CFT amplitudes are defined as
\eqn\hc{ A_n(b'_1, \ldots ,b'_n)
=  \left( \prod_{i=1}^n \int_0^{2\pi}\! d\phi_i~ \int_{r_+}^\infty \!
dr_i r_i \int_C \!
dt_i~  \right)
K(x_1,b'_1)\ldots K(x_n,b'_n) G_n(x_1, \ldots x_n)}
where the $n$-point Greens function $G_n$ represents the part of the
amplitude corresponding to bulk-bulk propagators only.  \hc\
corresponds
to nonderivative interactions, but the generalization is
straightforward.
The time integration contour $C$  runs down along the imaginary axis
from
$0$ to $-i\beta$.  As before we use time translation invariance to
shift
the contour in the real direction by $-T$, where we eventually take
$T \rightarrow \infty$.

Proceeding as in our other  examples, we want to continue $t'_i$
from the contour $C$ to either $t'_i=$real or $t'_i=$ real
$-i\beta/2$. Using the fact that all singularities are located as
in \ha,  the contour should be deformed as in Figs. 2 and 3.   The
region of integration along the real time axis corresponds to
$1_{++}$.  Continuing the the coordinate in $1_{++}$ by
$-i\beta/2$ takes us to region $1_{+-}$, so the second horizontal
time contour represents an integration of this region.  The two
vertical segments of the contour establish a correlation between
states in the two regions.  The entangled state is as in \ce,
\eqn\hd{ |\Psi \rangle = {1 \over \sqrt{{\cal Z}}}  \sum_n
e^{-\beta E_n/2} |n\rangle \otimes |n\rangle.}
By the same argument as in the Minkowski/Rindler example, this state
is equivalent to the one defined by a path integral on the lower half
portion of the Euclidean black hole ---
the Hartle-Hawking vacuum.  We again remark that the
fact that interaction vertices are to be included on the vertical
segments of the contour ensures that the energy eigenstates
appearing in \hd\ are those of the full interacting theory.

If we imagine first doing the integration over the vertical
segments then this leaves us with correlated boundary conditions
for the horizontal segments at large positive and negative BTZ
time.  In particular, it gives boundary conditions along the past
and future horizons in regions $1_{++}$ and $1_{+-}$.  Since $t =
+\infty$ corresponds to the future horizon in $1_{++}$ and the
past horizon in $1_{+-}$, boundary conditions on these two
horizons are correlated by the rightmost vertical segment.  And
similarly for the leftmost vertical segment.  The correlated
boundary conditions are equivalent to computing expectation values
in the state \hd.

Starting from  Euclidean propagators expressed in terms of
Euclidean time $\tau$, the arguments of the propagator can be
taken to either $1_{++}$ or $1_{+-}$ by the replacements
\eqn\he{ \tau ~ \rightarrow ~ \left\{ \eqalign{ &
e^{i(\pi/2-\epsilon)} t   \quad\quad ~~~~~~~~~~1_{++} \cr
 & e^{-i(\pi/2-\epsilon)}t +\beta /2 \quad \quad 1_{+-} } \right. }
For instance, the bulk-boundary propagator with both arguments in
$1_{++}$ is
\eqn\hf{\eqalign{&K^{(1_{++}1_{++})}(x,b') \cr &~~~~~~~~~
 = \sum_{n=-\infty}^\infty {1\over
\left[-\sqrt{{r^2-r_+^2 \over r_+^2}}\cosh(r_+ \Delta t)+{r\over
r_+}\cosh r_+(\Delta \phi+2\pi n)+i\epsilon \Delta t \sinh (r_+
\Delta t) \right]^{2h_+}} }}
Propagators with arguments in distinct regions do not need an
\eps\ prescription, since such propagators are nonsingular due to
the spacelike separation of points in $1_{++}$ and $1_{+-}$.

The Lorentzian prescription obtained by analytic continuation in
BTZ time is therefore to integrate vertices over regions $1_{++}$
and $1_{+-}$ with propagators given by the rule \he.  Furthermore,
we should also integrate over the the imaginary time segments
shown in Figs. 2 and 3, or equivalently impose correlated boundary
conditions on the horizons bounding the two regions.  This
prescription has also appeared in the recent work \Herzog.

With this prescription, the regions of the BTZ spacetime near the
singularities do not appear in the computation, and so it is clear
that there are no divergences from infinite blueshifts.  All
knowledge about physics in other regions besides $1_{++}$ and
$1_{+-}$ is contained in the Hartle-Hawking wavefunction.

This approach gives a satisfactory description involving only
regions $1_{++}$ and $1_{+-}$, but it is natural to expect that
there will exist alternative descriptions in which other regions
of the BTZ spacetime play a role.  Here an analogy with our
Minkowski spacetime example is helpful.  We saw that we would
analytically continue with respect to either Rindler or Minkowski
time.  In the Rindler case, which is analogous to using BTZ
coordinates, only the left and right wedges appeared in the final
result.   On the other hand, the full spacetime appears in the
Minkowski case, and so we would now like to find the analogous
continuation for the BTZ spacetime.  This is achieved by working
in Kruskal coordinates, as we now discuss.

\subsec{ Kruskal coordinates}

Lorentzian Kruskal  coordinates are defined as
\eqn\ia{ \eqalign{&x_1 = { 1+X ^2-T ^2 \over 1 -X^2
+T^2}\cosh{(r_+\phi)}, \cr &x_2 = { 1+X^2-T^2 \over 1 -X^2
+T^2}\sinh{(r_+\phi)}, \cr & x_3 = {2 X \over 1-X^2 +T^2} \cr &
x_0 = {2 T \over 1-X^2 +T^2}~. \cr}}
Note that $x_1^2 - x_2^2 \geq 0$, so given  \ad,  the coordinates
do not cover the regions 3 containing the closed timelike curves.
They do cover all of regions 1 and 2.  More precisely, they cover
all of regions 1 and 2 displayed in Fig. 1, but not those obtained
by periodically extending the figures in the vertical direction.
The AdS boundaries are at $X^2-T^2=1$, and we  approach either the
boundaries of $1_{+\pm}$ or $1_{\pm -}$ depending on whether we
approach $X^2-T^2 -1 =0$ from negative or positive values.  The
BTZ singularities are located at $X^2-T^2= -1$. The metric is
\eqn\ib{ds^2 = {4 \over (1-X^2+T^2)^2} \left[ -dT^2 + dX^2 +{r_+^2
\over 4} (1+X^2-T^2)^2d\phi^2 \right].}
For reference, the relation with BTZ coordinates in $1_{++}$ is
\eqn\bc{ r={1+X^2-T^2 \over 1-X^2+T^2}r_+, \quad \cosh{(r_+ t)}
={X \over \sqrt{X^2 -T^2}}, \quad \phi = \phi.}

The Euclidean signature  metric is
\eqn\ibb{ds^2 = {4 \over (1-X^2-\tau^2)^2} \left[ d\tau^2 + dX^2
+{r_+^2 \over 4} (1+X^2+\tau^2)^2d\phi^2 \right].}
 The
Euclidean manifold is given by the region $0 \leq X^2+\tau^2 \leq
1$. This metric is nonsingular since the proper length  of the
$\phi$ orbit cannot shrink to zero.  The metric near where the
denominator vanishes is that of
  AdS  in Euclidean Poincar\'e coordinates.
The boundary of the space is $X^2 +\tau^2 =1$, giving a torus.

Euclidean AdS/CFT amplitudes are now obtained by integrating
vertices over the Euclidean manifold.  However, analytic
continuation to Lorentzian signature is somewhat inconvenient
because of the constraint $0 \leq X^2+\tau^2 \leq 1$ on the
integration domain.
Since the range of the $X$ integration depends on $\tau$, one
finds a complicated analytic structure for the $\tau$ integrand.
Instead, it would be much more convenient if we could extend the
domain to the full $(X,\tau)$ plane.  This can be achieved as
follows.

We first observe that the metric is invariant under the {\it antipodal}
map defined
as \break $x \rightarrow x_A = -x$ where $x=(x_1,x_2,x_3,x_4)$.
{}From \ia\ with $T=-i\tau$ we see that in Kruskal coordinates the
antipodal map becomes
\eqn\ibc{X \rightarrow  {X \over X^2+ \tau^2}, \quad \tau
\rightarrow {\tau\over X^2 + \tau^2}.}
It follows that the
region $X^2 + \tau^2 \geq 1$ describes a second copy of Euclidean
BTZ, so if we extend our integration domain to the full $(X,\tau)$
cylinder we will be integrating over two copies of Euclidean BTZ.  It
is convenient to do this, and then divide by an appropriate factor
at the end of the calculation.

To see how this works in more detail, we first observe that under the
antipodal map \ibc\ Euclidean propagators transform as $G \rightarrow
(-1)^{2h_+}G$,  where the phase depends on how we choose to go
around the branch cut.   For example this transformation law
follows immediately for the Euclidean bulk-boundary
propagator from its form
\eqn\ie{K(x,b')=\sum_{n=-\infty}^\infty {(1-X^2-\tau^2)^{2h_+}
\over
 \left[ 2 XX' + 2 \tau \tau'
 -(1+X^2+\tau^2)\cosh r_+(\Delta\phi+2\pi
 n)\right]^{2h_+}}.}
This same transformation law holds for bulk-bulk Euclidean propagators
\refs{\Burgess, \Inami}.
Therefore, the effect of extending the integration with respect to
a given vertex to an integration over the full $(X,\tau)$ plane is
to multiply the original result by the coefficient,
\eqn\if{ 1 + \prod_i (-1)^{2h_{+,i}}}
where the product over $i$ is a product over propagators attached
to the vertex in question.  To reproduce the original result,
we should divide by the factor \if\ after extending each integration
to the two copies of Euclidean BTZ. In the supergravity limit, in
which we are working in this paper, $\sum_i 2h_{+,i}$ for
is always an integer, and the factor \if\
is either 2 or 0. If it is 2, we just have to multiply the
factor $1/2$ to each vertex after integrating it over the two
copies. On the other hand, if the factor \if\ is zero,
it means that the contributions from
the two copies cancel with each other. The method of doubling
the integration region is then not simply applicable in such a case,
and
a subtler analysis is required.   Of course for many reasons it would
be desirable
to find a way to carry out the analytic continuation
directly
for a single copy of the Euclidean BTZ with the constraint
$X^2 + \tau^2 \leq 1$.
In the following, we will consider the case when $\sum_i 2h_{+,i}$
is an even integer.

Now we proceed to analytically continue the Kruskal time arguments
of our Euclidean amplitudes.  The first step, as always, is to
locate the singularities in the complex $T$ plane.  There are two
kinds of singularities: those from the BTZ singularity and those
from lightlike separation.  The BTZ singularities are located on
the real $T$ axis at $T = \pm \sqrt{1+X^2}$. Lightlike
singularities in a propagator $G(x,x')$ occur when the geodesic
distance vanishes, $\sigma^2(x,x')=0$.   Examining the geodesic
distance, $\sigma^2 = (\Delta x_0)^2+(\Delta x_1)^2- (\Delta
x_2)^2- (\Delta x_3)^2~$ in the coordinates \ia, we find that with
$T'$ on the imaginary axis there are two singularities in the
complex $T$ plane, to the left and right of the imaginary $T$
axis.  Therefore, before doing any analytic continuation, the
singularity structure is as in Fig.  (6a).  Now when we continue
$T'$ to the real axis, the singularities also migrate to the real
$T$ axis.   The contour deformation is therefore similar to that
in Minkowski space with Minkowski time, and we obtain the contour
in Fig. (6b).   The novel feature is that the continuation tells us
how to integrate over both the BTZ singularities as well as the
usual lightcone singularities.
\fig{Integration contours in Kruskal time plane.  In the left hand
figure,
singularities on the real axis are due to the BTZ singularity; those
off
the real axis are lightcone singularities.
}{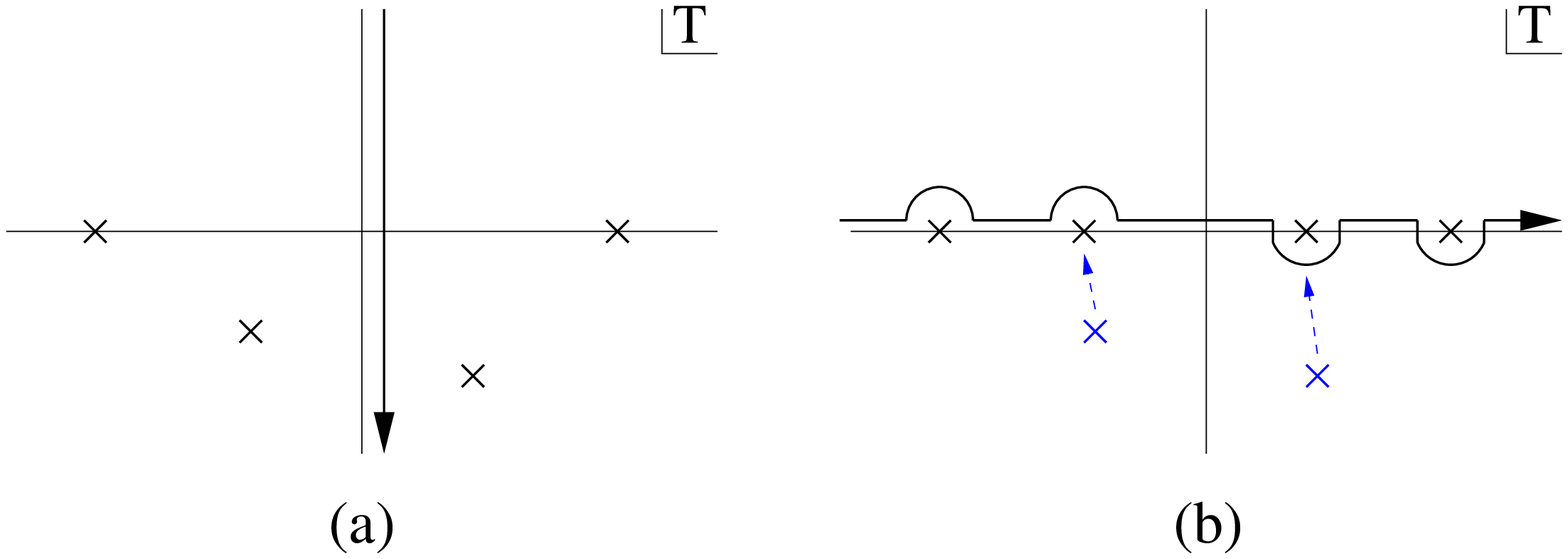}{4.5truein}
Our final result is that we are to integrate over all of regions 1
and 2 of Lorentzian BTZ in Fig 1, corresponding to the full $(X,T)$
plane.
with propagators obtained from Euclidean signature by the
substitution $\tau = e^{i(\pi/2 -\epsilon)} T$.  For instance, the
Lorentzian bulk-boundary propagator is
\eqn\ig{K(x,b')=\sum_{n=-\infty}^\infty
{(1-X^2+T^2-i\epsilon)^{2h_+} \over
 \left[ 2 X X' - 2(1-i\epsilon)TT'  -(1+X^2-T^2+i\epsilon)\cosh
r_+(\Delta\phi+2\pi
 n)\right]^{2h_+}}.}

Note that the integral domain is over the eight regions in the
left side of Fig. 1 --- four regions between the past
and the future singularities, and four more beyond the
future singularity. We also have to remember
that, since we started with two copies of Euclidean BTZ related
to each other by the antipodal map \ibc,  we needed
to divide the amplitude by the factor $2$. (We are assuming
that $\sum_i 2h_{+,i}$ is an even integer.)

The antipodal transformation,
\eqn\antipodallorentz{
 X \rightarrow {X\over X^2 - T^2},~~ T \rightarrow
 {T\over X^2 - T^2},}
maps the regions $1_{--}$ and $1_{-+}$, which are
beyong the future singularity, to the regions $1_{++}$
and $1_{+-}$. Under this map, the propagator
transforms as $G \rightarrow (-1)^{2h_+} G$.
Therefore,  rather than integrating over all the four
regions $1_{\pm \pm}$, we can restrict the integral to
the two regions $1_{++}$ and $1_{+-}$ and multiply the factor $2$.
This cancels the factor $1/2$ we introduced earlier to
extend the integration to the double of Euclidean BTZ. Thus
the net result is that we integrate over the regions
$1_{++}$ and $1_{+-}$ with the standard propagators
as in \ig . This result is reasonable since the boundaries
of these two regions are identified with the $(1+1)$ dimensional
spaces for the boundary CFT at finite temperature, as
discussed in section 5 \MaldacenaKR . If
the regions $1_{--}$ and $1_{-+}$ were included,
we would have had to
impose boundary conditions for these regions and the
question would have arisen whether there are
additional boundary CFT's for these.

The situation is more subtle when the integral runs over
regions of type 2. The antipodal transformation maps
$2_{--}$ and $2_{-+}$ to $2_{++}$ and $2_{+-}$ respectively.
Under this, the propagator transforms as
$G \rightarrow (-1)^{2h_+}G^*$, where the complex
conjugation means that we are using the opposite
of the standard $i\epsilon$ prescription. For the bulk-boundary
propagator, we can see this directly by acting \antipodallorentz\
on \ig , but it is also true for the boundary-boundary propagator.
Thus, if we want to restrict the integral region to be
over $2_{++}$ and $2_{+-}$, which are
between the past and the future singularities in
Fig. 1, we need to average over the two
opposite $i\epsilon$ prescriptions in an appropriate way. We will see
that this is closely related to the cancellation
of divergences at these singularities.

To summarize, the analytic continuation to the Lorentzian
BTZ using the Kruskal coordinates shows that
amplitudes are expressed in terms of integrals of
interaction points over the regions $1$ and $2$ between
the past and the future singularities.
For propagators in the region $1$, we use the standard
$i\epsilon$ prescription. On the other hand, for propagators
ending in the region $2$, we need to take an appropriate
average over signs of $i\epsilon$.

 \subsec{Integrating over the singularities}

The divergence of the propagator at the BTZ singularity has been
rendered finite by the \eps\ prescription, since
$1+X^2-T^2+i\epsilon$ is nonvanishing on the real $T$ axis.
Instead of the divergent behavior \am, we now have near the
singularities:
\eqn\ih{K~\sim~ f(X,T)\ln (1+X^2-T^2+i\epsilon)  ~ \sim ~ f(\Delta
t) \ln (r+\pm i\epsilon).}
The sign of \eps\ appearing in the last term depends on from which
BTZ region we approach the singularity.  A  naive \eps\
prescription would consist of adding a small imaginary part to BTZ
time and using the resulting propagator to integrate near the
singularity. This procedure leads to the divergent propagator of
\am\ and to divergent amplitudes upon integration over the
singularity.   But now we see that the correct \eps\ prescription,
written in terms of BTZ coordinates, adds an imaginary part to
both $r$ and $t$. Adding an imaginary part to $r$ lets us define
the amplitudes by integrating around the singularities in the
complex plane.  Analytic continuation has also been used
previously (though not derived from a consistent starting point)
in the context of quantum field theory near cosmological
singularities, e.g., \Tolley.

Let's examine the integration over the singularities in more
detail.  There are two BTZ singularities --- past and future with
respect to $1_{++}$ and $1_{+-}$
--- located at $T = \pm \sqrt{1+X^2}$.
Expressed in terms of BTZ coordinates, the metric
near either of the singularities is
\eqn\ii{ds^2 = r_+^2 dt^2 - {dr^2 \over r_+^2} + r^2 d\phi^2}
and the propagators behave as in \ih.  Examining \bc, we see that,
since our integration should extend over both sides of the
singularities (if we do not identify the integration regions using
the antipodal map), in  BTZ coordinates we should integrate over both
positive and negative $r$.   Positive and negative $r$ correspond
to the past and future cones of the Milne universe.  Note that we
do not integrate over the left and right cones of Milne, since
these correspond to regions of type 3, and these are not covered
by the Kruskal coordinates.

 We first consider the future
singularity. Approaching the singularity from $2_{++}$ we have the
relation (compare \af\ and \ia)
\eqn\ij{ r ~ \sim ~ {1+X^2-T^2 \over 2}.}
Therefore, propagators will diverge as $\ln (r+i\epsilon)$.  If we
take a generic derivative interaction, then the integration of a
vertex near the singularity will include a piece
\eqn\ik{ \int_{-r_c}^{r_c} \! dr\, {\ln^p (r+ i\epsilon) \over
(r+i\epsilon)^q}}
where $r_c$ is the radius where the propagators start to differ
from their leading behavior.  As $\epsilon \rightarrow 0$, \ik\
gives a finite,  but generically complex, result.

It is important that the imaginary parts arising from integration
over the two singularities combine in a manner consistent with
Hermiticity in the boundary CFT.  Without checking this explicitly
it is clear that this must come about, since our bulk amplitude is
mathematically equivalent to the analytic continuation of the
boundary CFT amplitude.   But to illustrate the point we can make
a simple check.
Consider a boundary correlation function for Hermitian operators
${\cal O}_i(0,\phi)$ evaluated at $t=0$ on the boundary cylinder.
Since the boundary theory is a tensor product, these operators can
be associated with either of the CFTs defined on $1_{++}$ or
$1_{+-}$.   Such a correlation function should be real, since all
operators are spacelike separated and hence commute.  When we
compute the amplitude in the bulk we pick up imaginary parts from
integrating over the BTZ singularities.  But due to the relation
$G \rightarrow (-1)^{2h_+} G^*$ under the antipodal transformation
in the region $2$, the imaginary parts cancel between the
singularities,
and the result is purely real as expected.

We have found that correlation functions computed in the BTZ black
hole are free from divergences and unphysical imaginary parts
because of the cancellation of effects at the past and future
singularities. This non-local cancellation mechanism may seem
surprising since it contradicts naive intuition that says that
points closer to a singularity should feel much more of its
effect. More quantitatively, in flat space a correlator falls like
some power of the distance and so if the two singularities are far
away the interaction points near the past singularity should make
a much smaller effect than the ones near the future singularity.
What makes a difference here is the asymptotically AdS boundary
condition of the BTZ black hole, which lets geodesics reflect off
the boundary and be refocused on future points. This makes it
impossible to effectively separate the two singularities.

\subsec{Defining scattering through the singularity}

An extremely interesting question concerns the
existence and behavior of scattering amplitudes for processes where
particles ``pass through" the
singularity.  This is the situation studied in
\refs{\BalasubramanianRY,\NekrasovKF,\SimonMA,
\LiuFT,\ElitzurRT,\CornalbaNV,\CrapsII,\FabingerKR, \LawrenceAJ,
\HorowitzMW, \Ben}.  The conclusion
of this work, \LiuFT\ in particular,
is that such scattering amplitudes are badly behaved in string
perturbation theory.

We might suppose that we could study such phenomena using the
techniques discussed earlier.
In particular we could study BTZ amplitudes with operators on the
boundary of regions $1_{-+}$
and $1_{--}$ as well as $1_{++}$ and $1_{+-}$.  Any particle path
between operators on
boundaries above and below the singularity will have to pass
through the singularity.

Formally we can calculate amplitudes like this by analytic continuation
\HemmingKD.  From \ae\
we see that we can ``move" an operator from region $1_{++}$
to $1_{-+}$ by analytically continuing in $\phi$,  much as in \bc :
\eqn\onemin{ \phi ~ \rightarrow ~ \left\{ \eqalign{ &[0,2\pi]
\quad\quad ~~~~~~~~~ 1_{++} \cr  &  [0,2 \pi] - i\beta/2
\quad\quad 1_{-+} } \right. }
The same continuation moves an operator
from region $1_{+-}$ to region $1_{--}$.

As we argued earlier, because the amplitudes we discuss are analytic in
$t$ and $\phi$
we do not expect singular
behavior for generic operator locations on the boundary of $1_{-+}$ or
$1_{--}$.
This seems to lead to
a conflict with the singular behavior found in the references cited
above.
It also conflicts with a naive assumption that an analytic continuation
for interaction point
integrations on a purely real Lorentzian slice of the BTZ space as
in \S 7
exists for such boundary operator locations.  If this were the case the
$i \epsilon$ singularities
would pinch
the contour at the BTZ singularity and make the integrated amplitudes
infinite in general.

One possible resolution concerns
the boundary conformal field theory we might expect to find on the
boundary of
$1_{-+}$ or  $1_{--}$.
The angular momentum operator that generates translations of $\phi$ has
spectrum unbounded
above and below.  So the sum over conformal field theory states is at
best conditionally convergent.
This suggests that correlation functions might not be derivable
directly from an operator formalism.
But this does not resolve the above conflict because the analytically
continued amplitudes might
well define a consistent bulk theory by themselves, without a boundary
field theory interpretation.

We believe the resolution to this problem lies in an obstruction to
performing the analytic continuation
of a boundary point into region $1_{-+}$ or $1_{--}$ with physical
contours for the interaction points.
We do not have a proof that
such an obstruction always exists but all our attempts have encountered
the same general problem.

This problem is illustrated by the following example.
We work in BTZ coordinates
and try to continue points to both
$1_{++}$ and $1_{-+}$.
$1_{++}$ corresponds to real t and real
$\phi$;  $1_{-+}$ corresponds to real t and $\phi = {\rm real} -
i \beta/2$.  Now, start from the Euclidean contour and first
continue to the real t axis for all points.   We can take the
contour to have three segments:  1) go from $-T$ to $+T$ along the
real
axis,
avoiding the singularities in the way which gives the Feynman
propagator;
2)  go from $+T$ to $-T$ along the real axis and underneath the
singularities;   3)  go from $-T$ to $-T - i\beta$.
Now we would like
to continue some of the boundary points to $\phi = {\rm real }
-i\beta/2$
while also moving the $\phi$  contour for segment (2) down by
$-i\beta/2$.  This cannot be done since the $\phi$ contour is pinched.
In particular, with the time argument given by segment (2),
singularities along this segment occur at
\eqn\a{ -\sqrt{{r^2 -r_+^2 \over r_+^2}} \cosh r_{+}(\Delta t -
i\epsilon)
+ {r \over r_+} \cosh r_{+}(\Delta \phi + 2\pi n)=0}
Expanding out the first $\cosh$ to first order in $\epsilon$, we
see that the imaginary part of $\Delta \phi$ changes sign depending on
the sign of $\Delta t$.   In general, both signs of $\Delta t$ occur,
so we will find  singularities just above the real $ \phi$ axis and
just below --- the contour is pinched.   This prevents us from moving
the
$\phi$ contour
downwards, unless we ``drag'' along some extra segment attached to
the singularities.   Other attempts result in the same pinching of the
$\phi$ contour.

This obstruction prevents us from obtaining a
simple picture of the Lorentzian signature amplitudes as integrals
over the interaction point locations on a real section of the
complexified BTZ space.
This removes the conflict with other approaches that study that
formulate the problem
on this purely real section.   But it also means that the techniques we
have developed
do not as yet resolve the issues raised in previous work.

\subsec{Remarks}

We have seen that a fixed Feynman diagram for correlators of
boundary operators in the BTZ geometry can be understood in
two different ways. First, as a Feynman diagram in which the locations
of the interaction vertices
are restricted to the regions outside the horizon.  This is the
``Rindler"
type description.  Second, as a diagram in which the locations are
integrated
over the full region covered by Kruskal coordinates, including regions
behind
the horizon and on both sides of the singularities.  This the
``Minkowski"
type description.  This identification suggests that certain things
about
physics behind the horizon can be learned from data located outside
the horizon. This idea is reminiscent of black hole complementarity.

In the second description, we integrate over interaction points
inside of the horizon as well as outside. Divergences and unphysical
imaginary parts, which could have appeared from an integral near
a singularity (and which do appear in similar computations in the Milne
universe\Ben),
are cancelled between the past and future singularities, at least in
one case.
At first glance this appears to be disturbingly nonlocal.  But the
singularities
of  eternal AdS-Schwarschild black holes are never extremely far apart.
Their
maximum separation is of order the AdS radius, no matter how large the
mass.
The shortest distance simple boundary correlators can resolve is also
AdS scale.
To observe the isolated, uncancelled singular behavior of one
singularity we
would have to use probes sensitive to  local bulk physics.   We expect
local correlators
of bulk supergravity fields to show such singular behavior\foot{Of
course such quantities are
not gauge invariant, but they may well be illustrative.  In the $2+1$
BTZ situation the
simplicity of the geometry allows cancellations to occur even for bulk
correlators.  This
follows from the antipodal symmetry of bulk-bulk propagators.}.
Extracting such local bulk physics from the boundary theory is a
notoriously difficult problem.
Perhaps the very complicated boundary operators necessary to localize
quantities in the bulk
will allow the well behaved boundary theory to display apparently
singular bulk behavior.

The factor \if\ that each Feynman diagram acquires
under Kruskal analytic continuation starting with
two copies of Euclidean BTZ is a major shortcoming of our approach.
In the supergravity limit, the factor is either $2$ or $0$
for each interaction vertex,
and we were able to find a way to perform the analytic
continuation in the Kruskal coordinates when it is $2$.
More generally, the factor is a complex-valued function
of mass. The factor cancels out if the interaction point
is in the region $1$, but it gives rise to a combination
of $G$ and $G^*$ with complex coefficients in the region $2$.
The mass dependence of these coefficients makes it difficult
to perform the analytic continuation in the full string
theory, though in that case we also need to discuss effects
due to twisted sectors, $etc$. It is desirable to find a
way to perform the analytic continuation starting with
a single copy of Euclidean BTZ.

Our conclusions do not lean heavily on being in three spacetime
dimensions, and one could extend our arguments to AdS black holes
in other dimensions.  Actually, much of what we say --- minus the
CFT interpretation --- could also be said for the four dimensional
Schwarzschild solution.  Green's functions defined in Euclidean
signature can be analytically continued to Lorentzian signature,
and in Kruskal coordinates will naturally lead to an integration
over the black hole singularities.\foot{A path integral
representation for the propagator was discussed from this point of
view in \HartleTP.}  One difference is that the Schwarzschild
solution is only an approximate solution of string theory, and so
the accuracy of the analytic continuation procedure needs more
careful justification.

In conclusion, our works illustrates the power of using analytic
continuation to define otherwise divergent Lorentzian amplitudes,
displaying a complementary correspondence between inside and outside
the horizon
phenomena in the process.

\bigskip\medskip\noindent
{\bf Acknowledgements:}

P.K. was supported in part by NSF grant PHY-0099590, H. O. was
supported
in part by
DE-FG03-92ER40701, and  S. S. was supported in
part by
NSF grant PHY-9870115.
We thank Micha
Berkooz, Ben Craps, Robbert Dijkgraaf, David Kutasov, Juan
Maldacena, Don Marolf, Emil
Martinec, Will
McElgin, Greg Moore, Rob Myers, and Lenny Susskind for discussions,
and
the Aspen
Center for Physics for hospitality during the initial stages of
this work.

\listrefs

\end